# Machines, AI, and the Past//Future of Things

**Karola Köpferl**
Chemnitz University of Technology
Faculty of Social Sciences
Junior Professorship of Sociology with Specialization in Technology
Chemnitz, Germany
karola.koepferfl@informatik.tu-chemnitz.de

**Albrecht Kurze**
Chemnitz University of Technology
Faculty of Computer Science
Chair Media Informatics

Chemnitz, Germany
albrecht.kurze@informatik.tu-chemnitz.de

**Abstract**
This essay explores a techno-artistic experiment that reanimates a 1980s East German typewriter using a contemporary AI language model. The project questions dominant narratives of progress by embedding generative AI in an obsolete, tactile interface. Through public exhibitions and aesthetic intervention, we demonstrate how slowness, friction, and material render artificial intelligence not only visible but open to critical inquiry. Drawing on concepts such as zombie media, technostalgia, and speculative design, we argue that reappropriating outdated technologies enables new forms of critical engagement. Erika—the AI-enabled typewriter—functions as both interface and interruption, making space for reflection, irony, and cultural memory. In a moment of accelerated digital abstraction, projects like this foreground the value of deliberate slowness, experiential materiality, and historical depth. We conclude by advocating for a historicist design sensibility that challenges presentism and reorients human-machine interaction toward alternative, perceived futures.

# Introduction

How does our relationship with technology change when artificial intelligence speaks through machines long deemed obsolete? This essay recounts a techno-artistic experiment that connects an East German GDR-era typewriter with a contemporary language model, opening up unexpected questions around trust, materiality, and the future of communication. At the intersection of human-computer interaction and critical media studies, the project demonstrates how hacking can serve as a method of critical engagement—rendering Artificial Intelligence not just visible but materially experiential. We trace how this interaction unfolds and what it reveals about our entanglements with machines, both past and future.

Our argument: the appropriation, misuse, and reinterpretation of outdated technologies is more than an artistic strategy. It is a way to question dominant narratives of technological progress, to re-materialize digital processes, and to provoke public discourse around the ethics and societal implications of AI. From the printing press to neural networks, the history of information technologies is one of ever-accelerating transformation. Large Language Models (LLMs) like ChatGPT now generate fluent text from minimal prompts in seconds—yet for most, their workings remain opaque (Heaven 2023; Perlman 2022).

# The Speed of Thought and the Slowness of Keys

The rise of generative AI has fundamentally transformed how we write, think, and communicate—often producing outputs faster than users can deliberate. Interfaces have become frictionless, minimal, spectral. In this environment, it is easy to forget that these systems have mechanisms, constraints, and cultural histories. They speak, but invisibly.

Our project set out to slow down these processes—not to resist technological development, but to encounter it differently. We wanted to feel AI working, not just consume its output. So we connected a 1980s East German typewriter—Erika S3004—to a large language model. The result was a conversation machine in which each word emerged with delay, rhythm, and mechanical echo. Erika made thinking tangible (Köpferl & Kurze 2024). The machine itself was never meant for this. Designed by VEB Robotron/Optima in the GDR in the mid-1980s, Erika was a bureaucratic tool: equipped with a daisy wheel, memory function, and correction key. Practical, efficient, unsentimental. And yet, decades later, it became the voice of a neural network.

The transformation was as much media archaeology as microelectronics. Using the typewriter's original serial port—once intended for printer use—we routed typed prompts through a WiFi-enabled ESP32 microcontroller to language models like ChatGPT and Mistral. The responses were printed back onto the same paper, character by character (Wahl 2023). There is no screen and no delete or backspace functionality. What emerged is a hybrid machine: the mechanical body of late socialism animated by a post-digital ghost.

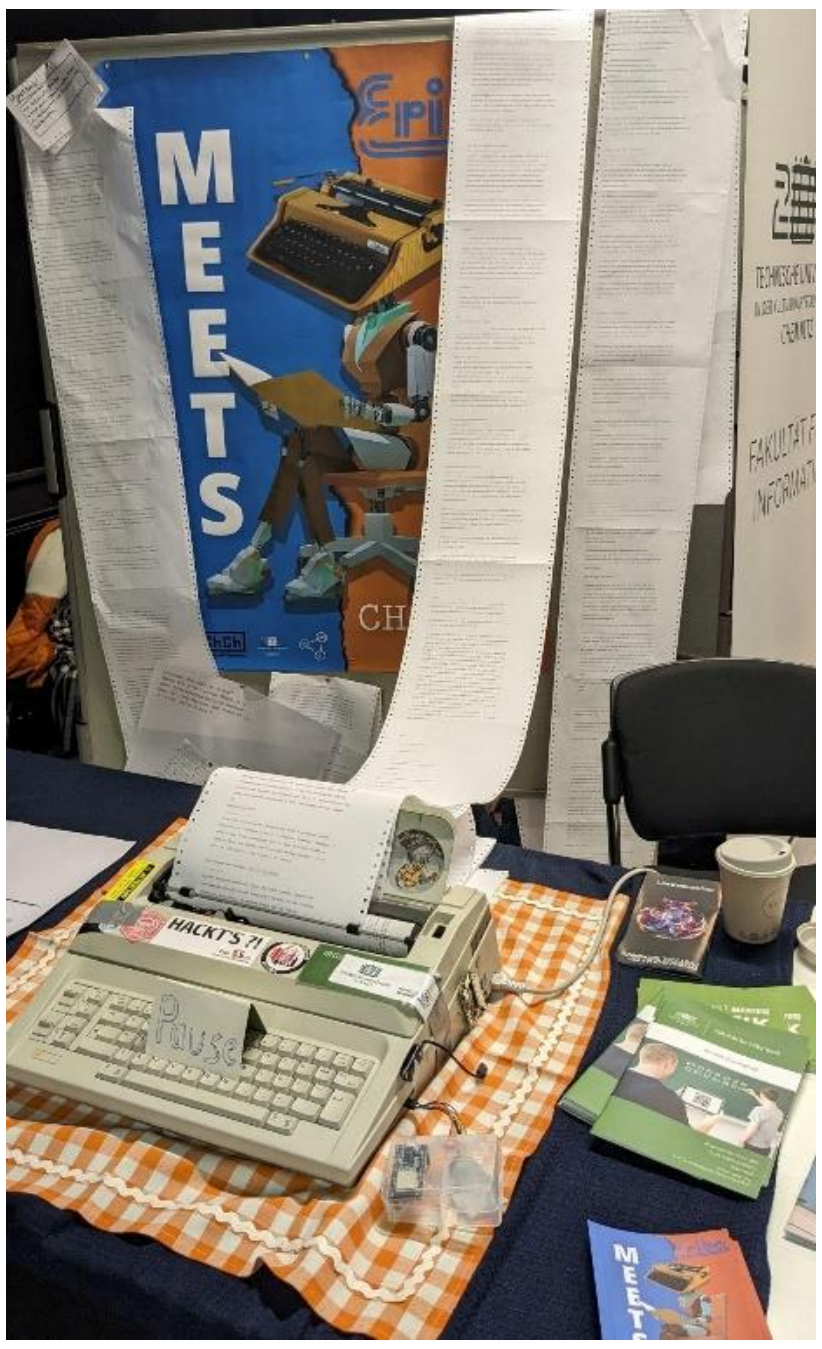

Figure 1: Erika connected to a LLM

This retrofit demanded more than clever wiring. Erika's 8-bit character set had to be decoded, mapped to Unicode, and reversed on output. Each input is limited to a single line. Each output arrives with the audible cadence of ribbon and ink—each letter a physical trace, not a collection of pixels. In public exhibitions, over 1,200 people have already sat down with Erika. Some had never touched a typewriter before; others hadn't in decades. They laid their fingers on the plastic keys of a device produced before German reunification and asked questions to an AI system hosted remotely. Many expected novelty. What they found was an aura—both familiar and strange. Mechanical choreography. Unexpected emotion. Some typed slowly, savoring the clicks. Others probed the AI with philosophical or playful questions: "Are you a ghost?" "What is the meaning of life?" "Can machines dream?" They waited minutes for a response—watching the answer unfold as if summoned, not computed. One visitor captured the mood best: "You can hear it think" (Köpferl & Kurze 2024).

The experience is not just aesthetic. By reintroducing latency and sound, the Erika typewriter renders the invisible logic of AI audible, tangible, and interruptible. Without screens or touch interfaces, there was no scrolling, no multitasking—only presence. This presence was not just attention—it emerged from the friction of delayed feedback, the physical accumulation of paper, and the palpable weight of each interaction. Unlike ephemeral digital interfaces where data vanishes from view, Erika materialized memory and demanded selective, deliberate engagement. The machine reclaimed space for doubt, curiosity, and irony. It invited people not only to ask what the machine says, but how and why it speaks at all. In this sense, Erika became a conversation piece in both meanings of the term: a medium for dialogue, and a catalyst for discussion. It refused the seamlessness of contemporary optimization and exposed instead a textured, resistant surface—a surface we could hear, touch, and wait for.

By slowing down the fast, we made it visible.

# Hacking as Method: Questioning Technology

At the core of the Erika project lies a deliberate act of misalignment. A typewriter was never meant to respond. Yet through wires and code, Erika became the physical voice of a disembodied AI. This tension—between form and function, past and future—is not a glitch. It is a principle. In this context, hacking becomes a mode of inquiry. Not simply the modification of hardware or writing of code but an epistemological stance: to make systems open—technically, culturally, and conceptually. To ask: What happens when machines are taken out of context, repurposed, made to perform outside their original design?

This approach draws from traditions in critical making (Ratto 2011), speculative design (Dunne & Raby 2013), and media archaeology (Parikka 2012). As Soro et al. (2019) argue, we must not only design the future but also "design the past"—treating history itself as malleable design material. Erika exemplifies this ethos: hacking becomes a means to resurface forgotten functionalities, to create friction, and to situate the digital within new historical trajectories.

In this sense, Erika is not only an interface anymore—it is a deliberate irritation.

Hertz and Parikka (2012) offer the concept of zombie media: technologies resurrected not to return nostalgically to their former use, but to haunt the present with unresolved tensions. These machines are not restored—they are repurposed as means of critique. In this framework, soldering wires, decoding obsolete character sets, and rerouting I/O becomes theory production through circuitry.

At exhibitions, Erika sparked curiosity not only about AI but about how it works, and how it can be implemented differently. Visitors peeked inside, asked about the firmware, and speculated on building their own. In that moment, hacking functioned as both installation and invitation—a transparent counterpoint to increasingly opaque AI infrastructures.

But Erika is not the limit of this approach. Once the principle of retrofitting obsolete machines with AI became tangible, new speculative questions emerged: What other devices might we reanimate?

Imagine a rotary phone—not wired to a switchboard, but to a synthetic voice model. You dial, and instead of hearing a friend, you talk to an AI that answers in the cadence of old conversations (Pollux Labs 2024). It replies with fragments of past dialogues, or hallucinated memories assembled from training data. This is not nostalgia—it is a simulation of that feeling. Or consider a VCR that plays back scenes from a past that never happened: AI-generated footage based on historically plausible prompts. The screen glows with manufactured memory, somewhere between retro-futurism and deep-fake melancholy. Or a Polaroid camera that prints instant photos of non-events—images imagined by a machine, triggered by keywords rather than light.

These speculative machines operate on a double register. They are playful and uncanny, comforting and disorienting. Their physical interface—a dial, a tape, a button—grounds the interaction in the past. But the intelligence behind it floats free, trained on datasets that abstract fragments of history without preserving historical continuity. Users are caught between presence and illusion, between authentic tactility and synthetic response. These fictions are not about machines.

They are about us:

- How do we perceive time?
- Whose memories are we reconstructing?
- What do we trust when everything feels real?

Such hybrids extend the Erika project's core questions: What happens when AI becomes slow, noisy, material—and uncanny? They allow us to explore how trust, temporality, and embodiment are reshaped by interfaces that blur memory and fiction.

# Technostalgia as Critique

At first glance, nostalgia appears to be the enemy of innovation—a sentimental retreat into the past that risks idealization and stasis. Our project suggests a different reading. In the case of Erika, nostalgia did not restore the old—it disrupted the new. Campopiano (2014:77) defines technostalgia as a sentimental or aesthetic attachment to obsolete technologies. As Campopiano (2014:76) observes, technostalgia—our fondness for outdated technology—is not necessarily regressive. It can be a critical tool. When past machines are reanimated with present functions, they do not merely return—they interfere. They unsettle. This interference not only disrupts contemporary AI but also reconfigures our understanding of the original artifact, altering its cultural meanings through integration into new practices. They show us what we might have overlooked in the sleekness of the now. Rather than conceal AI behind smooth UX, these speculative machines reframe it. They make the future feelable—and ask whether we are ready to live with its ghosts.

But in our case, it became more than a feeling: it became a method. Technostalgia operated as a form of defamiliarization—a lens through which AI became strange, embodied, and open to inquiry. Erika was not designed to look retro. It simply was: loud, mechanical, resistant to the smoothness of contemporary interaction. And in that material friction, it produced irritation—and attention.

This unsettling effect is key to technostalgia's critical potential. When a neural network speaks through a GDR-era typewriter, the result is not just dialogue—it is temporal dissonance. A friction between planned economies and platform capitalism, between analogue bureaucracy and algorithmic logic. In such encounters, participants do not only ask what the machine says, but what it means that it says anything at all. In this sense, nostalgia becomes a vehicle for critique. Alizadeh et al. (2022:2) argue that outdated technologies, when reactivated, can prompt reflection and turn memory into "design material." Erika's form, its slowness, its noise—all became aesthetic signals that recontextualized AI, inviting speculation and irony instead of mere awe.

Visitors did not simply use Erika. They speculated with and about it. They asked:

- Are you haunted?
- Can machines dream?
- Do you remember?

These were not functional questions. They were ontological ones, triggered not by the intelligence of the model, but by the constraints and aura of the interface itself. As Odom et al. (2012:816) suggest, material friction can deepen engagement and foster critical awareness. Technostalgia, in this way, becomes an active disruption. It reframes AI not as seamless innovation but as contested terrain—where form matters, where history lingers. Erika's materiality, its GDR-industrial shell, represents more than retro aesthetics. It evokes histories of control, collectivity, and latency.

Soden et al. (2021:459) argue for a historicist sensibility—one that resists "presentism" and insists on contextualizing technology within longer arcs of change. Erika embodies this principle. It refuses the clean line of progress, insisting that the past is not dead but still present, folded into every keystroke. Nostalgia, as Dang et al. (2023:1000) note, is ambivalent. It can inhibit innovation when it becomes mere longing—but it can also promote it, by grounding speculative futures in collective memory. Erika enacts this tension. Its voice leans toward tomorrow, its body holds onto yesterday. In that contradiction lies its power.

Rather than idealize the past, technostalgia as critique renders the present strange. It poses a series of questions: What kind of machines do we want? What futures do we recall? And what must we remember, in order to imagine otherwise.

# Conclusion: The Next Decade of Things

As generative AI increasingly integrates into everyday life—hidden within seamless interfaces, background processes, and voice assistants—it risks becoming imperceptible. Its very ubiquity renders it opaque. Yet invisibility is not neutrality. The smoothness of AI conceals the social, historical, and material conditions of its emergence.

Projects like Erika disrupt this trajectory. They do not seek to replace the future with the past, but to intervene in the present—to make visible what has become hidden, and to open to critical inquiry what is taken for granted. By embedding AI in a slow, noisy, obsolete device, we reframed it not as a tool, but as a conversation. Not as convenience, but as provocation.

Soro et al. (2019) describe designing the past as an act of imaginative reappropriation: a way to use memory as material, and history as a horizon. In this sense, Erika is not merely retro. It is recursive. It allows us to revisit the evolving notion of intelligent machines with a different tempo, one keystroke at a time.

Nostalgia can both hinder and enable innovation. Its ambivalence is not a flaw but a latent opportunity. Erika activates that potential. In the click of its keys, in the hum of its motor, and in the permanence of ink on paper, we rediscover not only what we have left behind—but what remains worth carrying forward.

Perhaps the most radical interfaces will not be those that are most efficient, but rather those that slow us down, make us wait, and demand that we listen. Perhaps the next decade of "things" will not be defined by novelty, but by recognition.

# Acknowledgments

This research was supported by the Junior Professorship of Sociology with Specialization in Technology and the AI Lab of Chemnitz University of Technology. We also thank the Chaostreff Chemnitz and its members for their technical support and encouragement. In realizing this project, we built upon the work of the Open Source and maker community, whose resources provided the foundation for our technical implementation. In particular, we relied on community-digitized documentation of historical East German office and computing technologies, which enabled us to design a system suitable for continuous public use and adaptable to new large language models as they emerged.

# References

Alizadeh, F., Mniestri, A., Uhde, A., & Stevens, G. (2022). On appropriation and nostalgic reminiscence of technology. In Proceedings of the CHI Conference on Human Factors in Computing Systems Extended Abstracts (pp. 1–6). ACM. https://doi.org/10.1145/3491101.3519676

Campopiano, J. (2014). Memory, temporality, & manifestations of our technostalgia. Preservation Digital Technology & Culture, July 2014. https://doi.org/10.1515/pdtc-2014-0004

Dang, J., Sedikides, C., Wildschut, T., & Liu, L. (2024). More than a barrier: Nostalgia inhibits, but also promotes, favorable responses to innovative technology. Journal of Personality and Social Psychology, 126(6), 998–1018. https://doi.org/10.1037/pspa0000368

Dunne, A., & Raby, F. (2013). Speculative everything: Design, fiction, and social dreaming. MIT Press. https://doi.org/10.1093/jdh/epv001

Heaven, W. D. (2023). Artificial intelligence: ChatGPT is everywhere. Here’s where it came from. MIT Technology Review. https://www.technologyreview.com/2023/02/08/1068068/chatgpt-is-everywhere-heres-where-it-came-from/

Hertz, G., & Parikka, J. (2012). Zombie media: Circuit bending media archaeology into an art method. Leonardo, 45(5), 424–430. https://doi.org/10.1162/LEON_a_00438

Köpferl, K., & Kurze, A. (2024). Human-Chatbot Interaction: When ChatGPT meets an old Typewriter. In Proceedings of the 21st International Conference on Culture and Computer Science: from Humanism to Digital Humanities (KUI '24). ACM. https://doi.org/10.1145/3719236.3719250

Odom, W., Zimmerman, J., & Forlizzi, J. (2012). Material representations: Expanding interaction design research methods. In Proceedings of the Designing Interactive Systems Conference (DIS 2012) (pp. 599–608). ACM. https://doi.org/10.1145/2317956.2318044

Perlman, A. (2022). The implications of ChatGPT for legal services and society. SSRN Electronic Journal. https://doi.org/10.2139/ssrn.4294197

Pollux Labs. (2024). ChatGPT im Telefon – ein Retro-Sprachassistent. Pollux Labs. https://polluxlabs.net/raspberry-pi-projekte/chatgpt-im-telefon-ein-retro-sprachassistent

Ratto, M. (2011). Critical making: Conceptual and material studies in technology and social life. The Information Society, 27(4), 252–260. https://doi.org/10.1080/01972243.2011.583819

Soden, R., Ribes, D., Avle, S., & Sutherland, W. (2021). Time for historicism in CSCW: An invitation. Proceedings of the ACM on Human-Computer Interaction, 5(CSCW2), Article 459. https://doi.org/10.1145/3479603

Soro, A., Taylor, J. L., & Brereton, M. (2019). Designing the past. In Proceedings of the CHI Conference on Human Factors in Computing Systems Extended Abstracts (pp. 1–10). ACM. https://doi.org/10.1145/3290607.3310424

Wahl, D. (2023). KI-Projekt: ChatGPT auf einer Schreibmaschine mit ESP32 umsetzen. Heise Make Magazin, 2, 28–32. https://www.heise.de/ratgeber/KI-Projekt-ChatGPT-auf-einer-Schreibmaschine-mit-ESP32-umsetzen-7545547.html